\documentclass[]{piparticle}
\usepackage{graphicx}
\usepackage{amsmath}
\usepackage{animate}
\usepackage[colorlinks=true,allcolors=blue]{hyperref}

\begin{document}

\runningauthor{G Abramson}  

\title{Serendipitous observation of a coronal mass ejection during the total solar eclipse of 14 December 2020}

\author{G Abramson,\cite{inst1,inst2,inst3}\thanks{E-mail: abramson@cab.cnea.gov.ar}  }

\pipabstract{
We report observations of the total solar eclipse of 14 December 2020, during which a coronal mass ejection can be see	n propagating. A comprehensive set of photographs covering a high dynamic range of exposure allow to characterize its dimensions. The displacement of the front can be seen occurring during the few minutes of totality. 
}

\maketitle

\blfootnote{
\begin{theaffiliation}{99}
   \institution{inst1} Centro At\'omico Bariloche (CNEA), R8402AGP Ba\-ri\-lo\-che, Argentina.
   \institution{inst2} Instituto Balseiro, Universidad Nacional de Cuyo, R8402AGP Bariloche, Argentina.
   \institution{inst3} Consejo Nacional de Investigaciones Cient\'{\i}ficas y T\'ecnicas, Argentina.
\end{theaffiliation}
}

\section{Introduction}
Coronal mass ejections (CMEs) are violent expulsions of plasma, magnetic field and energy from the Sun~\cite{howard2006,alexander2006,howard2011}. During a typical CME millions of tons of solar material are ejected into the heliosphere, within a time scale of about 1 hour. The ejection consists of particles moving at velocities averaging several hundreds of kilometers per second. If adequately oriented, they can reach the Earth after a couple of days, disturbing the magnetosphere and the ionosphere, disrupting radio communications and putting space-borne electronics at risk.

They have been recognized as a unique solar phenomenon since the 1970's, particularly after their observation from space with the very sensitive coronographs on board Skylab, when they were called a variety of names before settling  on the current denomination~\cite{alexander2006}. Those transient events were soon recognized to be the cause of the high speed solar wind responsible for the geomagnetic storms long observed in radio frequencies, and indeed their effects had been observed on Earth for thousands of years in the form of polar auroras. After many years of study, and even though it is known that the coronal magnetic field is the driving agent, central questions about their cause and relevance in the dynamics of the corona still remain~\cite{howard2006,howard2011,koutchny2004}.

Shortly after their discovery from space-borne observatories, the record of solar observations was searched for possible CMEs, particularly during solar eclipses. Just one instance was found~\cite{eddy1974}: a drawing made by Gugliemo Tempel during the eclipse of 18 July 1860 in Spain shows a swirl of the size of the solar disk, with a bright core, in the midst of mainly radial streamers of the corona. Several other observers along the path of totality described and sketched a similar structure (but an equal number didn't, indicating the subtlety of the phenomenon to the naked eye) and, put together, their reports seem to correspond to the development of a CME~\cite{eddy1974}. Since the times of Tempel, the growing interest in solar phenomena has made that each eclipse was more observed than the previous ones, yet no other clear observation of a CME was recorded for a full century. Indeed (according to Alexander~\cite{alexander2006}) the rate of CMEs, their typical duration and the brief periods of totality combine themselves to allow about one chance per century of capturing one during an eclipse. Besides Tempel's remarkable sketch, Filippov et al.~\cite{filippov2020} mention a little known work showing a sketch (Fig. 18 of \cite{bugoslavskaia1950}) with a structure resembling the front of a light bulb shaped CME during the eclipse of 21 September 1941, and the seemingly corresponding eruptive prominence (ibid. Fig. 49). We have not been able to find photographs of this eclipse of the pre-CME recognized era which, occurring during World War II, was not widely observed. Less clear examples are known, resembling disconnected CMEs in the final stage of evolution~\cite{cliver1989}. 

Even though total solar eclipses allow a brief observation of coronal phenomena, the tight natural occultation provided by the Moon is ideal to study the inner corona, which is normally occulted by coronographs below 2--2.5 solar radii, thus hiding about 80\% of the mass of the K-corona and its dynamics~\cite{koutchny2004}. In such a context, both modes of observation of the corona, the continuous one from space borne coronographs and the sporadic one during eclipses, complement themselves. Indeed, on a few occasions the phenomena related to CMEs have been observed in such manner. Perhaps the most relevant is the one studied by Koutchny et al.~\cite{koutchny2004,koutchny2002}, who observed the eclipse of 11 August 1999 a few hours before a large CME was observed from space. They were able to identify the precursor of the CME during the eclipse. 

Very recently, Filippov et al.~\cite{filippov2020} reported the observation of traces of a very slow (250 km/s) CME during the eclipse of 21 August 2017, using high resolution and high dynamic range photography, with space-based instruments being able to follow it before and after. Likewise, during the eclipses of 13 November 2012 and 3 November 2013, Druckm\"uller et al.~\cite{druckmuller2017} were able to observe that the cores of previous CMEs had evolved into tethers connecting to tornado-type prominences, providing evidence for the flux of electrons from interplanetary traveling CMEs back to the Sun. 

In summary, total solar eclipses are excellent opportunities to observe the phenomena of the inner corona, complementary to those made with coronographs from ground and space. In this work we report observations made during the eclipse of 14 December 2020. Totality touched land only in the Northern Patagonia of Chile and Argentina. Taking place during the travel restrictions imposed by the COVID-19 pandemic, very few expeditions were able to make their way to the path of totality. Coincidentally, a large CME erupted during the eclipse, and its propagating front can be seen in the obtained photographs. From high dynamic range and edge enhanced images we were able to derive its main dimensions and velocity. 

\begin{figure*}[t] 
\begin{center}
\includegraphics[width=\textwidth]{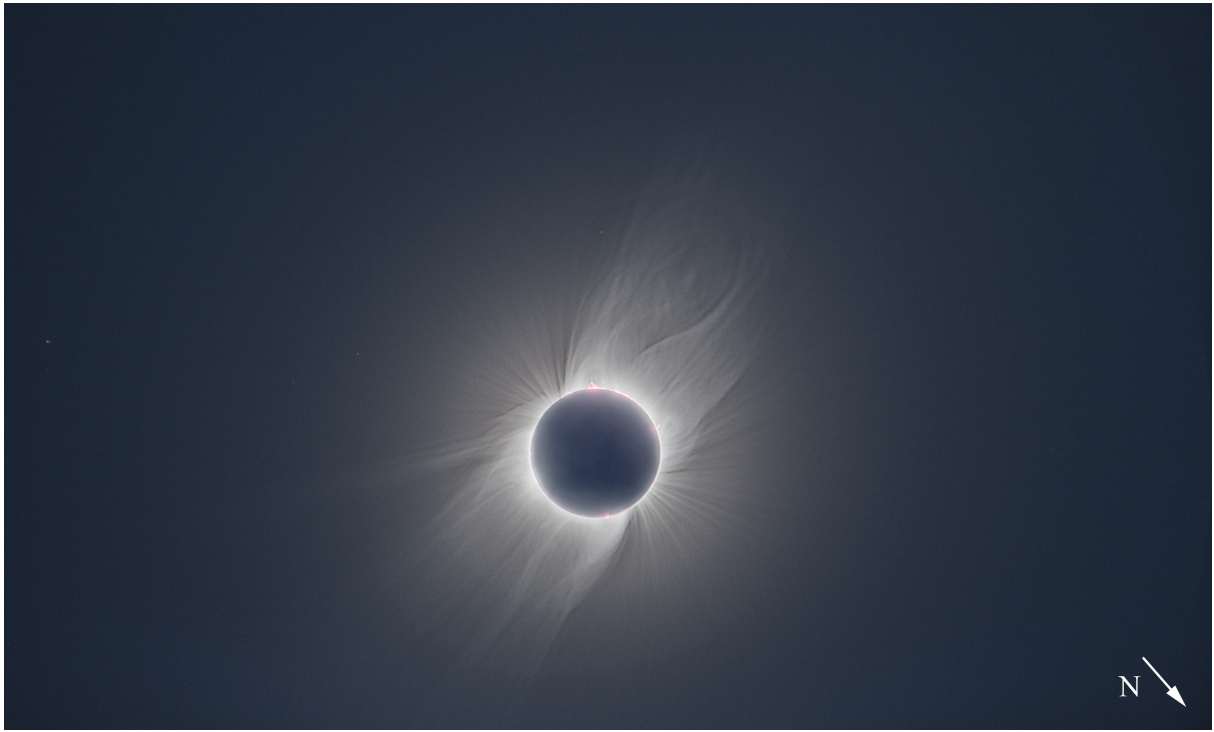}
\end{center}
\caption{High dynamic range composite corresponding to the first set of images taken during totality (from 16:08:10 to 16:08:21 UT). The teardrop shaped feature in the corona is the CME described in the text. In the full resolution image the comet C/2020 X3 can also be seen (see Supplementary material). The arrow marks solar north.} 
\label{fig:hdr}
\end{figure*}

\section{Observational materials and methods}

The eclipse was observed with portable equipment from an outback site near Piedra del \'Aguila, province of Neuqu\'en, Argentina. The exact location is latitude: $-40^\circ\,0'\,18.4''$, longitude: $-69^\circ\,59'\,6.3''$, elevation: 450 m, on the northern shore of Limay river, next to the Pichi Pic\'un Leuf\'u dam. At this site we enjoyed 122.6 s of totality, between 16:08:03.7 UT and 16:10:06.3 UT (using $\Delta T =69.4$ s and lunar limb correction,~\cite{jubier2020}), at an elevation of $72^\circ$. The sky was clear during most of the eclipse, with scattered cumulus clouds at the end of the second partial phase. Wind was very strong, sustained around 50 km/h from the west, with gale force gusts (not measured, based on own experience living in Patagonia, and according to forecast). The equipment was set up downwind from a very thick bush of \textit{Rosa eglanteria}, which provided partial shelter under its lee. 

The photographic setup consisted of a Canon T3i camera with a Tamron 18-270 mm F/3.5-6.3 Di II VC PZD zoom lens, set at the longest focal length. This provided a field of view of $6.3^\circ$ (diagonal), with a resolution of $3.6''$ per pixel. The camera was mounted on an iOptron SkyTracker camera mount, which provided sidereal tracking to help keeping the Sun in view. Since this is a very simple mount without fine controls, to allow fine adjustments during the eclipse it was mounted, in turn, over a manual equatorial mount and tripod (EQ-1 type). Polar alignment was performed with a magnetic compass, taking into account the magnetic declination of the site. Camera and mount were powered by a 12 V, 17 Ah gel battery.

Exposure control was programmed and run in-camera, via a Lua script run by the Magic Lantern alternative operating system available for Canon cameras~\cite{magiclantern} (see Supplementary material). The script ensured autonomous and automatic imaging during all the relevant phases of the eclipse, including rapid bursts of shooting during totality. Timing was controlled by the camera clock, synchronized in the morning via NTP to an error of less than one second. 

During the partial eclipse the lens was covered with a neutral density filter ($N\!D \approx 5$), custom made from unbranded polymer film provided by Saracco Astronom\'{\i}a (Buenos Aires). The image was focused manually on the solar limb, using the camera display and its provided magnification. After first contact, focus was checked using the lunar limb. The filter was removed approximately 15 seconds before totality, and replaced after third contact. The focus was not changed for the unfiltered exposures, since previous trials had deemed it unnecessary.

\subsection{Complete set of images}

The automatic capture script was programmed to shoot 7 sets of exposures during the partial phases of the eclipse, each set consisting of a bracket of 3 exposures. Immediately before third contact, 3 relatively long exposures were made to capture the popular ``diamond ring'' effect, followed by 7 short exposures in rapid succession to reveal the rugged lunar limb (the ``Baily's beads''). The same scheme run around fourth contact, reversed. During totality, the script run 8 sets of shots (7 of them were completed), of varying exposures, each covering a range of 11 photographic stops. At maximum eclipse, a 3 shots bracket of relatively long exposure was taken to capture ashen light and the outer corona. The maximum exposure was 2 seconds, which is about the limit to prevent trailing artifacts with the implemented optical setup. Table~\ref{fotos} summarizes the photographic parameters of the 155 shots obtained.

\begin{table}[t]
\caption{Complete set of photographs obtained during the eclipse. The eighth set of exposures taken during totality was incomplete at C3.}
\label{fotos}
\begin{tabular}{lllc}
\hline
 Aperture (F/) & Exposure (s) & ISO & Images  \\ \hline
 \multicolumn{4}{l}{Partial: C1-C2, C3-C4} \\ 
 11 & 1/50  & 100 & 7, 7 \\
 11 & 1/80  & 100 & 7, 7 \\
 11 & 1/125 & 100 & 7, 7 \\ \hline
 \multicolumn{4}{l}{Second and third contacts: C2, C3}  \\ 
 16 & 1/800 & 100 & 7, 7 \\ 
 16 & 1/60  & 400 & 3, 3 \\ \hline 
 \multicolumn{4}{l}{Totality: C2-C3} \\ 
  8 & 1/250 & 100 & 7 \\ 
  8 & 1/250 & 200 & 7 \\ 
  8 & 1/125 & 200 & 7 \\ 
  8 & 1/60  & 200 & 7 \\ 
  8 & 1/30  & 200 & 7 \\ 
  8 & 1/15  & 200 & 7 \\ 
  8 & 1/8   & 200 & 8 \\ 
  8 & 1/4   & 200 & 8 \\ 
  8 & 1/2   & 200 & 8 \\ 
  8 & 1     & 200 & 8 \\ 
  8 & 1     & 400 & 8 \\ 
  8 & 1     & 800 & 8 \\ \hline
 \multicolumn{4}{l}{Totality (maximum) } \\ 
  8 & 1/2   & 800 & 1 \\
  8 & 1     & 800 & 1 \\
  8 & 2     & 800 & 1 \\ \hline
\end{tabular}
\end{table}

\subsection{Scale of the photographic images}

For the purpose of taking measurements on the images, a precise scale was determined using the silhouette of the lunar limb during totality. Fifty points were placed at random around the limb and, from their coordinates, a least squares fit of a circle produced a radius $r_{m} = 275.4 \pm 0.8$~px (see Supplementary material). We used one of the images with shortest exposure to minimize the bleeding of the brightness of the lower corona over the lunar limb (which brightest parts were, nevertheless, avoided for the placement of points). Combined with a lunar apparent radius of $1000.45''$ (topocentric, derived from DE440~\cite{de440,chevalley2021}), we have a scale: 
\begin{align}
\sigma = 3.63 \pm 0.01~\text{arcsec/px}.
\end{align} 

\section{Results}
Figure~\ref{fig:hdr} shows a high dynamic range composite of the first set of images, at the beginning of totality. Several large prominences appear around the lunar limb. They could be seen clearly with the unaided eye or with small binoculars, contrasting sharply with a very bright and white lower corona. The tear-shaped feature that stands out in the corona is the front of the coronal mass ejection that occurred during the eclipse. The three main components of typical CMEs can be identified: a front, a cavity left behind, and a bright core within it. The phenomenon was not discernible visually during totality (at least, it was not noticed by us, but it may have been more apparent to others, as noticed by Eddy~\cite{eddy1974} regarding the eclipse of 1860), but it stood out immediately when checking out the success of the exposures in the field, after totality (see Supplementary material).

\begin{figure}[t] 
\begin{center}
\includegraphics[width=\columnwidth]{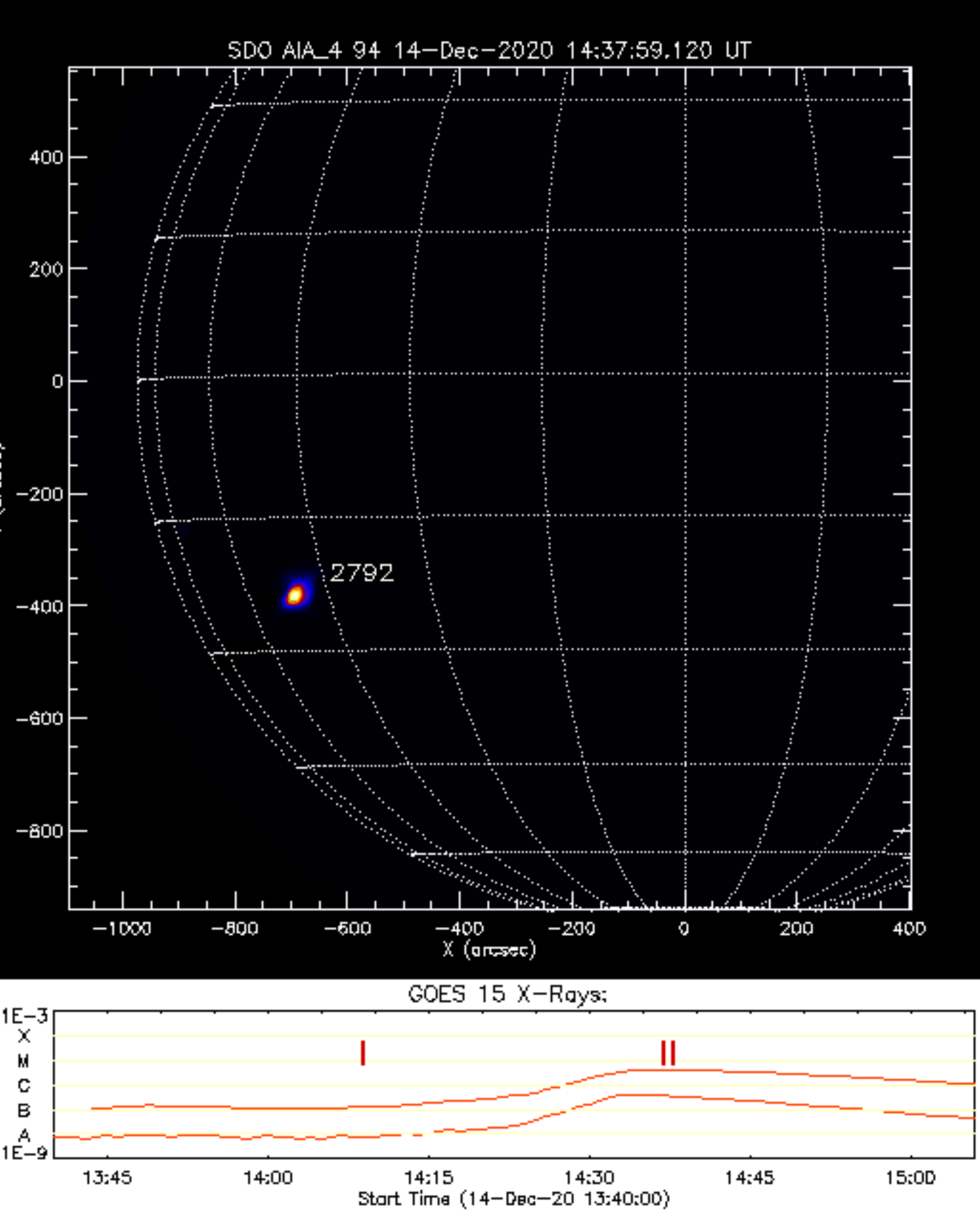}
\end{center}
\caption{Flare produced by active region 12792 during the eclipse, which produced the CME obseved during totality. SDO AIA 94 \r{A} image and GOES 15 X-rays data, from SolarMonitor~\cite{solarmonitor} (\url{https://www.lmsal.com/solarsoft/latest_events_archive/events_summary/2020/12/14/gev_20201214_1409/index.html}).} 
\label{fig:flare}
\end{figure}

The CME originated from active region 12792, a small bipolar region (area 10 millionths, Hale class $\beta$, McIntosh class Bxo) which at the time of the event presented four small spots. The region produced a single X-rays flare, GOES class C4.0, which developed between 14:09 UT and 14:56 UT, with a peak at 14:37 UT (8 minutes before first contact at the site of observation)\footnote{From SolarMonitor, hosted by the Solar Physics Group, Trinity College Dublin and the Dublin Institute for Advanced Studies \cite{solarmonitor}. Data can be accessed at \url{https://www.solarmonitor.org/index.php?region=12792&date=20201214}.}. Figure~\ref{fig:flare} shows the location of the eruptive event, identified at heliographic coordinates (S23,E50). 

\begin{figure}[t] 
\begin{center}
\includegraphics[width=0.7\columnwidth]{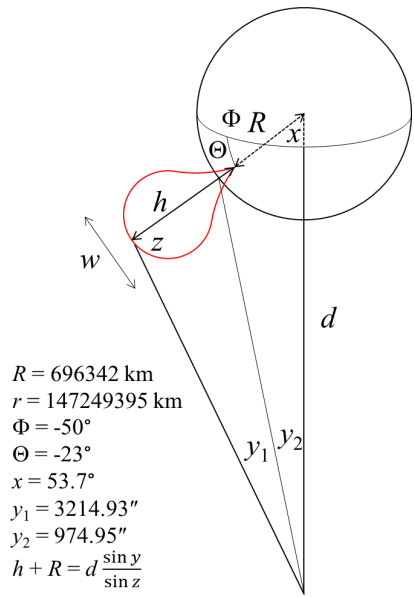}
\end{center}
\caption{Relevant magnitudes for the calculation of the true size of the CME. Parameters $d$ and $y_2$ are from ephemeris~\cite{de440,chevalley2021}, $R$ is the solar radius~\cite{solarradius}, $\Phi$ and $\Theta$ are the heliographic coordinates of the associated flare~\cite{solarmonitor} (from which the angle $x$ is calculated). Measured from our images are $y_1$ and the angle subtended by $w$. Angles $y=y_1+y_2$ and $x$ give the third angle of the triangle, $z$, and the law of sines allow to solve for the height of the CME, $h$.} 
\label{fig:angles}
\end{figure}

The dimensions of the CME can be determined from our images and solar ephemerides. Figure~\ref{fig:angles} shows the relevant magnitudes to be determined for the calculation. The angle $y_1$ subtended by the longitudinal extension of the CME was measured from the photographs, as well as the angle subtended by the transversal dimension of the CME front. The angle $y_2$ subtended by the solar radius, as well as solar distance $r$, were obtained from ephemerides (topocentric at the site of observation, ephemeris DE440~\cite{de440,chevalley2021}). The heliographic position of the flare allows to calculate the angle $x$, responsible for a strong perspective foreshortening of the CME as seen from Earth. Assuming a radial development of the CME, these parameters allow to solve the triangle and calculate the true height of the front of the CME over the solar surface, $h$. A similar calculation (to take into account their closer distance) gives the transverse width of the front, $w$, and the size of the prominent core seen inside the CME, $b$. The main source of error of these derived magnitudes is the imprecision in the determination of the position of the front, estimated at 10 px in the image. Combined with the calculated error of the photographic scale, and assuming that the other parameters have negligible errors compared with these, we have calculated an error of 26000 km in the following magnitudes:
\begin{center}
\begin{tabular}{cl}
\hline
dimension & $\times 10^6$ km \\  \hline
$h$ & $ 2.960 $ \\
$w$ & $ 1.052 $ \\
core & $ 0.185 $
\end{tabular}
\end{center}

\begin{figure*}[t] 
\begin{center}
\includegraphics[width=\textwidth]{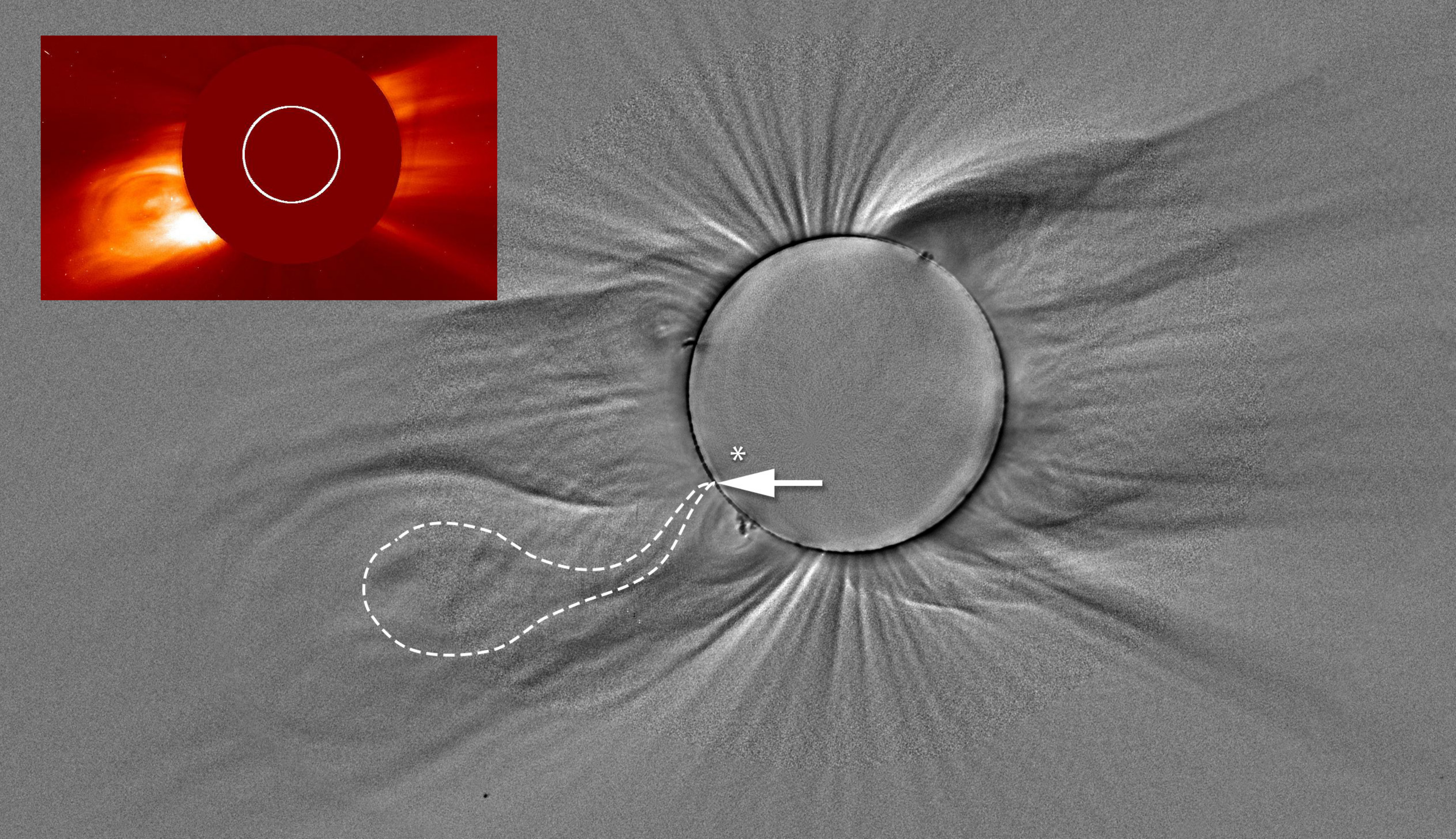}
\end{center}
\caption{Edge-enhanced view of the solar corona, showing the extension of the CME down to the limb. The asterisk marks the position of the flare. The dashed outline highlights the core and a structure that may connect it to the active region. The inset is SoHO's LASCO C2 image at the closest time (16:12 UT); the red circle is the shadow of the mask of the instrument, and a white circle represents the Sun. Solar north is up. Credit of the inset: ESA/NASA-SoHO/LASCO.} 
\label{fig:difs}
\end{figure*}

\begin{figure*}[t] 
\begin{center}
\includegraphics[width=\textwidth]{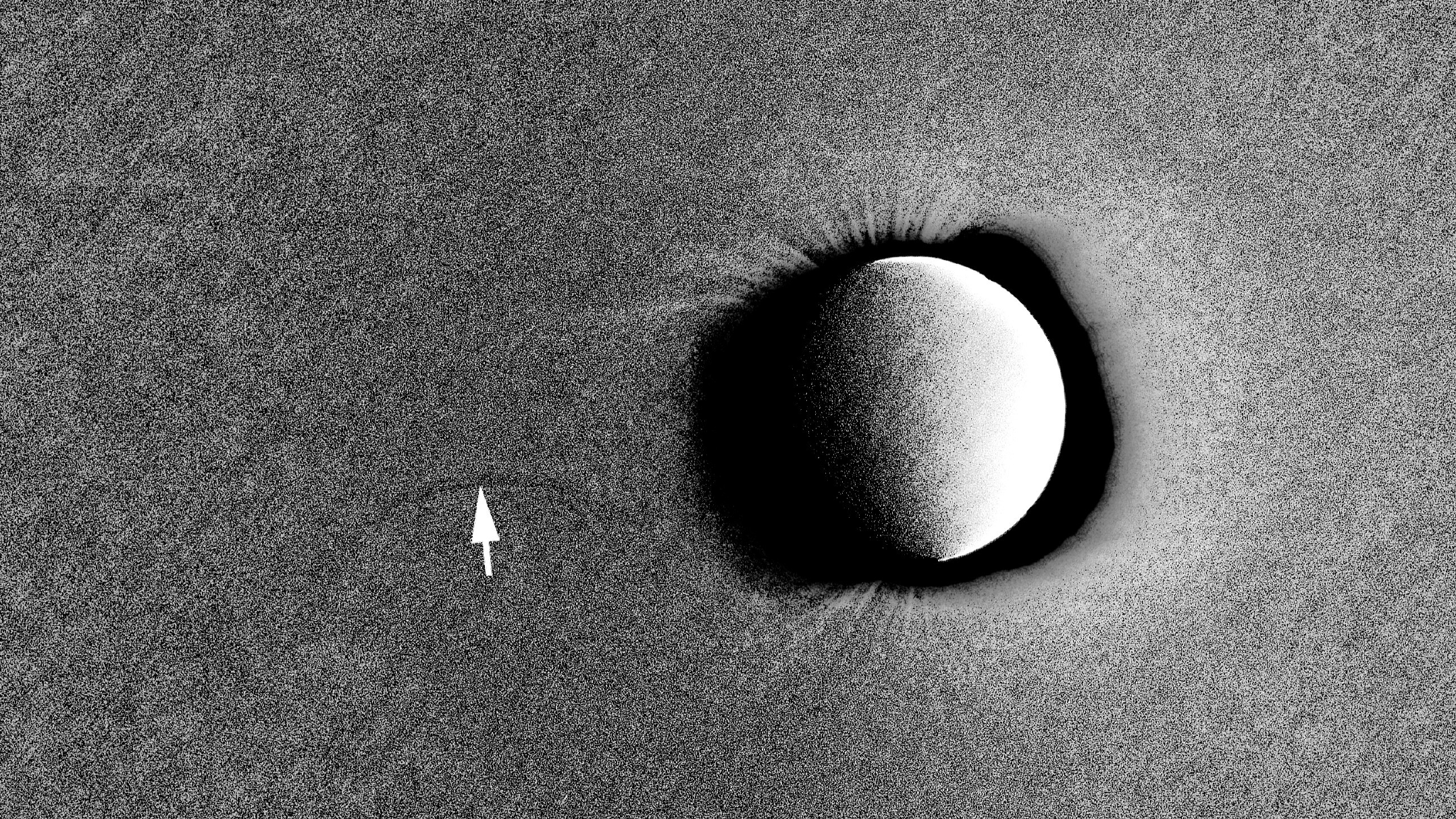}
\end{center}
\caption{Difference between images taken at 16:09:53 and 16:08:14 (F/8, 1/4 s, ISO 200), unprocessed to prevent distortions. The arrow shows the northern front of the CME, with the greatest displacement. Solar north is up.} 
\label{fig:movement}
\end{figure*}

Given the (peak) time of the flare and the time of the exposures, we know that the CME had expanded for 91 minutes. This corresponds to an average radial velocity of $542\pm 5$ km/s (to be compared with the plasma radial velocity of the solar wind, measured by the ACE satellite around 350 km/s at the moment of the event~\cite{solarmonitor}), and a transverse expansion of $108\pm 5$ km/s from the middle line. Figure~\ref{fig:difs} shows another view of the corona, produced an edge enhancement technique consisting in radially blurring the different exposures and subtracting them from longest to shortest. This view  shows that the CME-blown structure still extended to the lower corona, and it probably still anchors in the chromosphere at the place (marked with an asterisk) were it originated 91 minutes before. The dashed outline shows the bright core and a probable tether connecting it to the active region, in the manner observed in~\cite{druckmuller2017}. The inset is the closest image taken by the solar coronograph LASCO C2, on board the Solar \& Heliospheric Observatory (NASA/ESA), showing the corresponding structures in the corona. It emphasizes the relevant role of the observation during eclipses, which allows to observe regions of the solar corona usually masked out by coronographs. 

The estimated velocity, at the distance of the propagating front, lies within the limit of resolution of our images for the duration of totality. We made an attempt to measure it directly, with the result shown in Fig.~\ref{fig:movement}. This shows the difference between two corresponding images of the first and last sets of photographs, uunprocessed to avoid distortions that edge enhancement could introduce. They were registered using two of the stars visible in the field: 51 Oph and 44 Oph, while the brighter $\theta$ Oph was disregarded for being closer to the edge of the frame, where the field is more distorted. An animated version of the two images can be found in the  Supplementary material. The dark stripe signaled with an arrow shows that the northernmost part of the front shifts visibly during the 99 seconds interval of the two shots. From this feature we have estimated a width of 8--12 pixels, corresponding to an apparent angle of 29--64 arcseconds. At the distance of the core (which is closer than the Sun due to perspective, as estimated before) the displacement corresponds to 18--28$\times 10^3$ km, hence a velocity in the range of 182--283 km/s, compatible with the transversal expansion rate estimated above. Unfortunately, we were not able to perform a similar calculation for regions closer to the solar surface, sine the stars are not clearly visible in the shorter exposures. Other techniques, along the lines of the ones implemented in the very high resolution images processed by Druckm\"uller~\cite{druckmuller2017}, might be able to extract further information from this region.
 
\section{Conclusions}
The serendipitous observation of an ongoing CME, reported here, emphasizes the importance of careful observation of the solar corona during total solar eclipses. They allow to observe and record details of the corona down to the chromosphere,  normally hidden by the occulters of coronographs. We have shown that even the displacement of coronal features can be measured during the brief minutes of totality, using off-the-shelf photographic equipment. It is our belief that more results could be derived from the obtained set of images, especially in combination with similar records taken eastward from our location, where totality occurred up to 10 minutes later. We hope that a special issue like the present one might bring us into contact. 

\begin{acknowledgements}
Data supplied courtesy of SolarMonitor.org and NASA/ESA-SoHO. The author acknowledges partial support from the following grants: CONICET PIP 112-2017-0100008 CO, UNCUYO SIIP 06/C546, ANPCyT PICT-2018-01181, as well as the support of Gabriela Casella and Mar\'{\i}a Penovi during the expedition. 
\end{acknowledgements}

\newpage
\section*{Supplementary Material}
\setcounter{figure}{0}

\begin{figure}[h] 
\begin{center}
\includegraphics[width=\columnwidth]{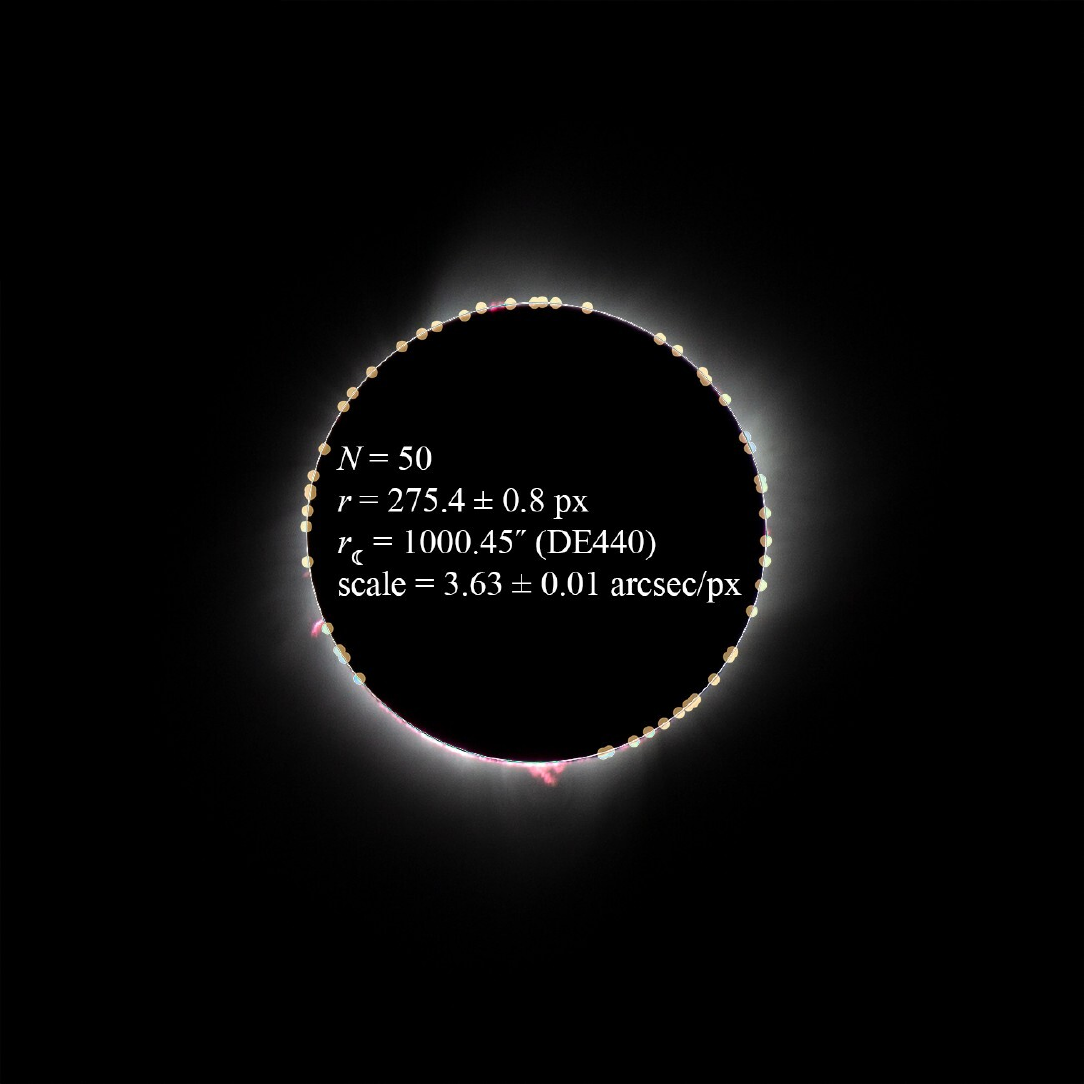}
\end{center}
\caption{Least squares fit of a circle to 50 points set over the lunar limb. The brightest parts were avoided for improved accuracy. The size of this circle was used to set the scale of the photographs.} 
\label{fig:cicle}
\end{figure}

\begin{figure}[h] 
\begin{center}
\includegraphics[width=\columnwidth]{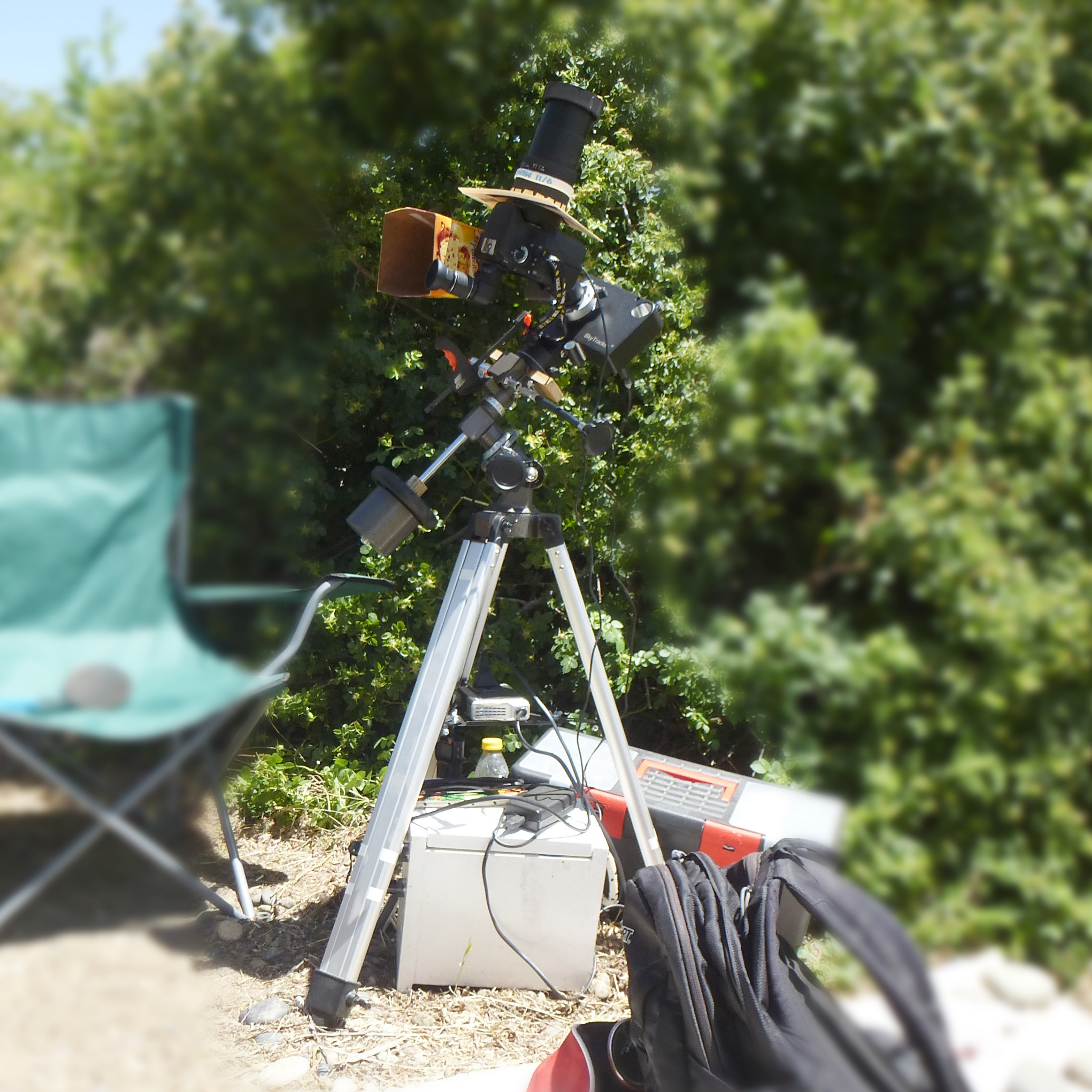}
\end{center}
\caption{Photographic rig used during the eclipse. Canon T3i camera, Tamron 18-270 mm zoom lens, iOptron SkyTracker camera mount, equatorial mount and tripod (EQ-1 type), 12 V gel battery. 
} 
\label{fig:rig}
\end{figure}

\begin{figure}[h] 
\begin{center}
\animategraphics[autoplay,loop,width=\columnwidth ,keepaspectratio]{1}{comp-}{0}{1}
\end{center}
\caption{The animated gif shows the two images used to produce Fig.~5. A large mask was used to cover the Moon, which moves significantly between the shots.} 
\label{fig:check}
\end{figure}

\begin{figure*}[t] 
\begin{center}
\includegraphics[width=\textwidth]{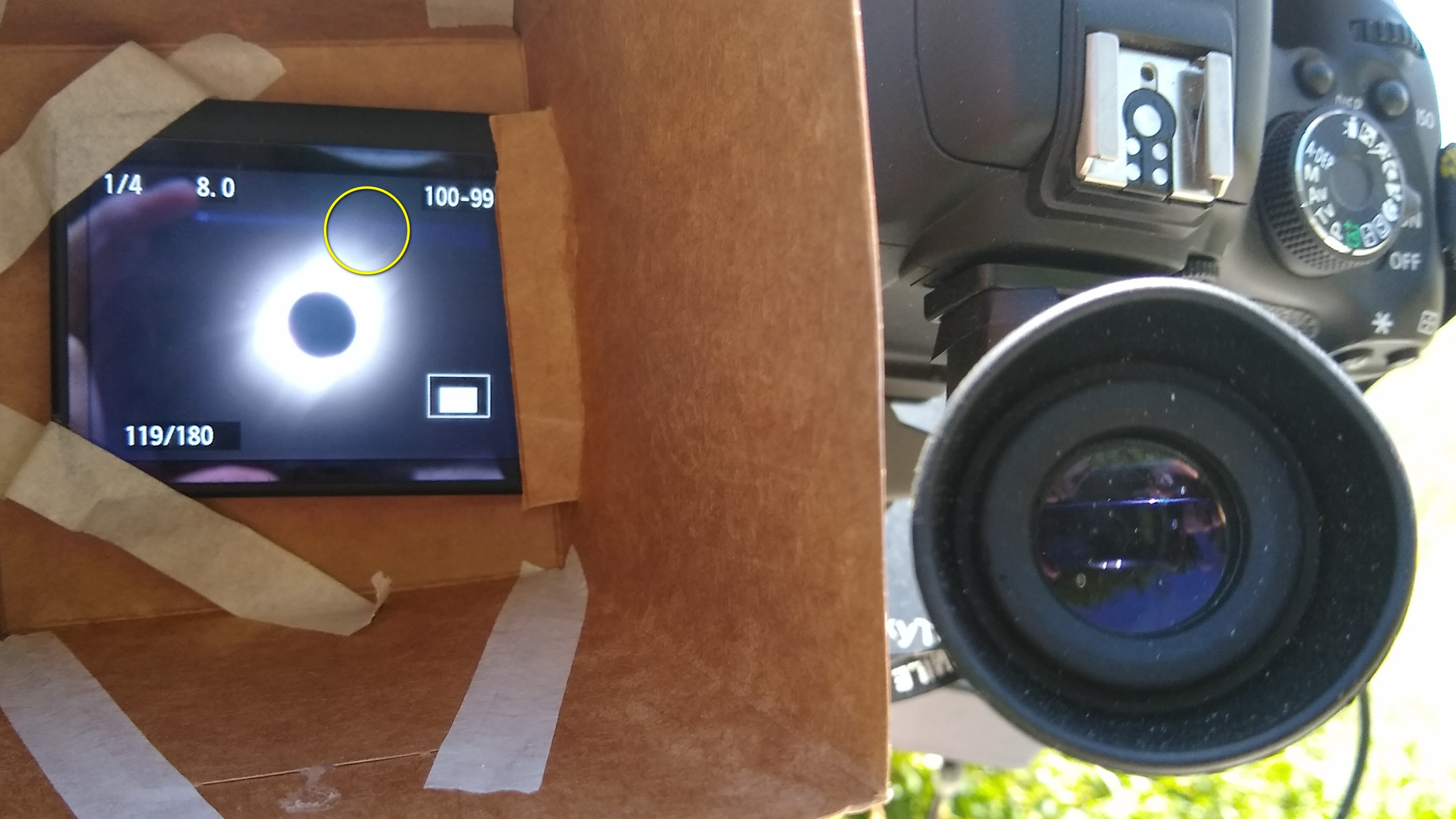}
\end{center}
\caption{A quick check in the field after the eclipse showed an unusual feature in the corona, which turned out to be the unexpected CME.} 
\label{fig:check}
\end{figure*}

\begin{figure*}[h] 
\begin{center}
\includegraphics[width=\textwidth]{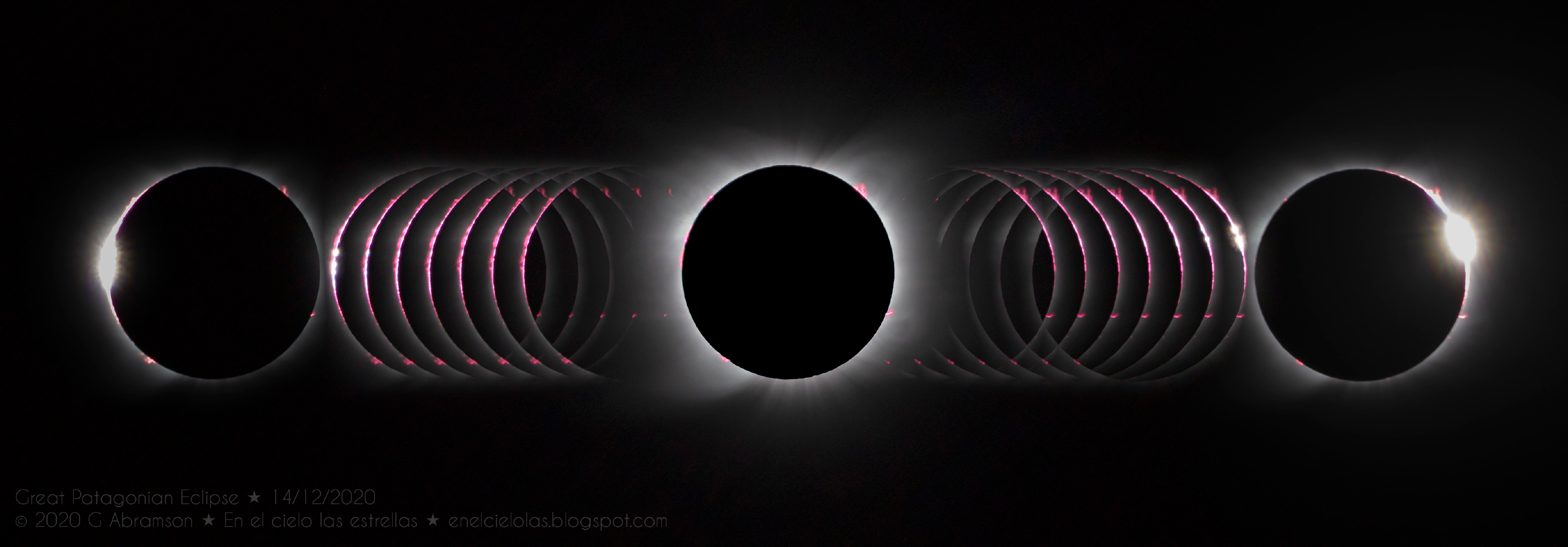}
\end{center}
\caption{The chromosphere was captured by two sets of fast shots at second and third contacts. } 
\label{fig:chromosphere}
\end{figure*}

\begin{figure*}[h] 
\begin{center}
\includegraphics[width=\textwidth]{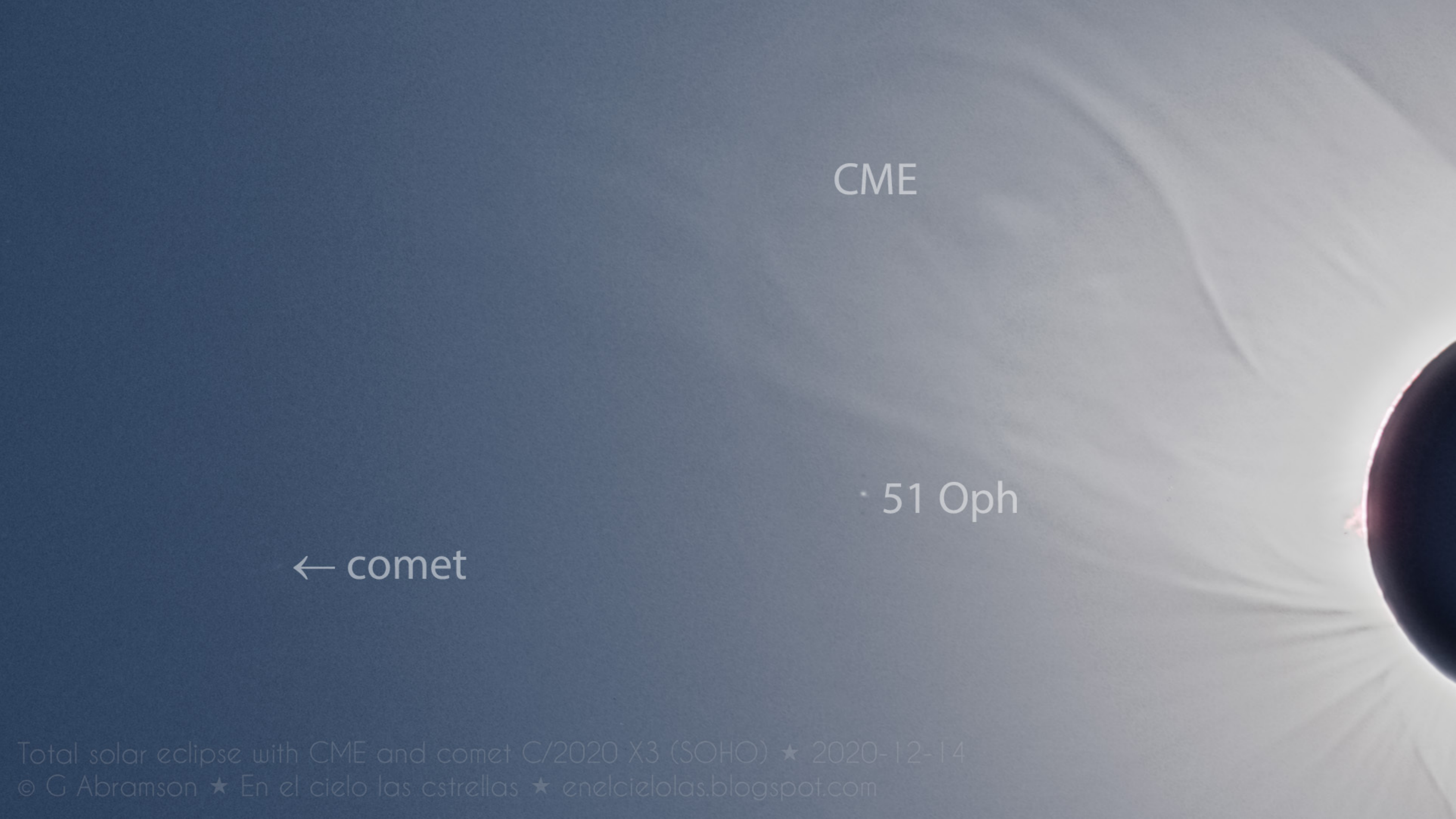}
\end{center}
\caption{Sungrazer comet C/2020 X3, discovered the previous day in SoHO images by Worachate Boonplod, was captured approaching the Sun. SoHO images show that the small comet did not survive perihelion.} 
\label{fig:comet}
\end{figure*}

\onecolumn
\section*{Camera control script}

This is the Lua script used to automatically take the photographs during the eclipse. It runs in-camera, making use of the Lua implementation of MagicLantern for Canon cameras, which is currently in the experimental builds: \url{https://builds.magiclantern.fm/experiments.html}. The script is based on the original one by Brian Greenberg. Besides recentering of the Sun (if needed), it runs unsupervised. If interrupted, it can be run at any moment and will take over from the current events based on the camera clock.

\tiny
\begin{verbatim}
-- Eclipse Magic 
-- Programmed exposure sequence for a solar eclipse
-- Copyright 2017 by Brian Greenberg, grnbrg@grnbrg.org.
-- Modified by G Abramson 2019-2020
-- Distributed under the GNU General Public License.
-- 1. Set TEST or SHOOT
-- 2. Set Contact times for test and shoot
-- 3. Set Exposure settings for all phases
-- 4. Check (set) camera time
-- 5. Run script from ML menu

require ("logger")

-- Variable definitons that have to go here.  Ignore them.
c1 = {}
c2 = {}
c3 = {}
c4 = {}

-- *** TEST: SET 1. SHOOT: SET 0. ***********************************************
-- Test: beeps and logs, no shoot, time starts counting with script run. 
-- Shoot: actually shoot, log, timed with camera clock
TestBeepNoShutter = 0

-- *** CONTACTS *****************************************************************
-- Set the 4 contact times here.  Time zone is irrelevant, as long as your camera and these
-- times are the same.  Make sure the times are correct for your location, and that your camera
-- is accurately set to GPS or NTP time.
--
if ( TestBeepNoShutter == 0 )
then
-- Piedra del Águila (playas) 70 00 31.91 W 40 03 16.16 S 
	c1.hr = 11; c1.min =  45; c1.sec = 40;
	c2.hr = 13; c2.min =  08; c2.sec = 05;
--  MAX     13	          09           05 
	c3.hr = 13; c3.min =  10; c3.sec = 02;
	c4.hr = 14; c4.min =  35; c4.sec = 50; 

else
-- Testing:
	c1.hr = 00; c1.min =  00; c1.sec = 30;
	c2.hr = 00; c2.min =  08; c2.sec = 07;
	c3.hr = 00; c3.min =  10; c3.sec = 01;
	c4.hr = 00; c4.min =  15; c4.sec = 30;
end

--
-- Send log information to file ECLIPSE.LOG at the
-- top directory of the camera card.  Useful for testing.
--
LogToFile = 1
LoggingFile = nil

-- APERTURE
-- Set an aperture value.  The script assumes that the aperture stays constant throughout the
-- eclipse.  The camera will (try to) set this aperture (f-number) at the beginning of the script.
-- Useful, as it is easy to forget this, if you are shooting with a regular camera lens.
-- If the script and camera are being used with a fixed aperture lens (or telescope), then
-- set the "SetAperture" to 0.
SetAperture = 1
Aperture    = 8    -- Aperture for totality
C23Aperture = 16   -- Smaller aperture for nicer ring and beads
PAperture   = 11   -- Aperture for partial (filtered)

--
-- If you shoot with LiveView active, you will reduce the amount of mirror slap, giving
-- less vibration.  But, some cameras (Such as the 5DmkII.) have a limited range of shutter 
-- speed when LV and Movie mode are enabled.  This reduces the available speeds from 30 seconds 
-- through 1/8000th of a second to 1/30th to 1/4000th of a second.  Exceeding that range while 
-- LV and Movie mode are enabled will crash the script.
--
-- Setting this variable to 1 will, before taking any images, check if LV is running, and if it
-- is, will tell you to turn it off.  If you have disabled movie mode (a menu option on some cameras,
-- like the 5DmkII, and a hardware switch on others) you should set this to 0.
--
-- Test your camera.  If the exposure speeds you want crash the script in LV, check how to disable
-- movie mode.
--
WarnLiveView = 0

--
-- If you're shooting in Live View (to reduce mirror slap) you'll probably still want to check
-- your focus occasionally.  However, the status display of the script pretty much fills the screen,
-- making this difficult.  Enabling HideConsole will display the console for the indicated number
-- of seconds both before and after the next shutter event, and turn the console off between.  This
-- also gives you an idea of how much time you have before the next shutter event -- if the console
-- is hidden, you have at least ConsoleShowDelay seconds to mess around.  If you turn this off, the
-- script will start with the console displayed, and you will have to manage it through the Magic
-- Lantern menu.  If you turn it off, it stays off until you turn it on again.
--
HideConsole = 1
ConsoleShowDelay = 30

--
-- Even if you shoot with Live View with Silent Mode 1 (which forces the mirror to be 
-- locked up, and eliminates the vibration from the shutter opening) there will be slight 
-- vibrations introduced as the shutter closes.  According to Jerry Lodriguss at
-- http://www.astropix.com/wp/2017/07/17/mirror-slap-and-shutter-shock/ these vibrations
-- are most prominent between 0.125s and 2s, and can be somewhat mitigated by a slight 
-- delay before exposure, to allow the vibration to dampen a bit.
--
-- Enabling this option lets you set a range of shutter speeds to delay before, and 
-- how long (in milliseconds) to pause.
--
DoShutterShockDelay   = 1
SlowestDelayedShutter = 2
FastestDelayedShutter = (1/8)
ShutterShockDelayMS   = 300		-- Value is in milliseconds!


-- ** EXPOSURE SETTINGS *****************************************************************
-- Partial phase settings. 
--
PartialISO            = 100
PartialShutterSpeed   = (1/80)  -- Filter Saracco OK
PartialMarginTime     = 30			-- Number of seconds after C1 or C3 and before C2 or C4 to start exposures
PartialExposureCount  = 72   		-- Number of partial phase exposures before and after totality
PartialDoBkt          = 1		  	-- Do you want to do exposure bracketing?  1 - yes, 0 - no
PartialBktStep        = 0.666667		-- Number of f-stops in each step.  Can 0.333333, 1, 2, etc
PartialBktCount       = 1				-- How many brackets on each side of the neutral exposure?

-- ** BAILY's BEADS ****************************************************************
-- Do a fast burst of exposures at C2 and C3, to try to get Baily's beads and chromosphere.
-- You will need to know how many exposures your camera will buffer, and how long it takes the
-- buffer to fill.  At "StartOffset" seconds before C2 or C3, the camera will take "BurstCount"
-- exposures, as fast as it can.  You should adjust C23StartOffset so that burst of image straddles 
-- the contact time.  Note that, between the setting of the camera clock and the jitter in this
-- script, there will be some error in the timing.  +/- half a second or more is possible.
--
C23BurstCount        = 7 	-- Note that most Canon DSLRs can't take more than 13-14 RAW images
								          -- in a burst before the buffer is full, and they slow to ~1 image/second.
C2BurstStartOffset   = 4
C2BurstTime          = 8
C3BurstStartOffset   = 4
C3BurstTime          = 8
C23BurstISO          = 100
C23BurstShutterSpeed = (1/800)


-- ** DIAMOND RING ********************************************************************
-- Do a fast burst of exposures before C2 and after C3, to try to get the diamond ring.
--
-- Be careful setting the RingStartOffset, Count and Time.  If the burst of images for the
-- pre-C2 rings runs longer than expected, it can cause the pre-C2 Baily's Beads exposures to
-- be skipped.
--
-- The post-C3 Rings exposures (if enabled) will run immediately after the post-C3 Baily's Beads
-- exposures.
--
DoRing          = 1		-- Are we going to try for a burst for the diamond ring?
RingStartOffset = C2BurstStartOffset + 5  -- How long before C2/after C3 to start?  Be careful that
										 -- this does not interfere with the Baily's burst!
RingBurstCount  = 3  -- How many images?
RingBurstTime   = 4  -- If we're doing a manual burst, how long should it last?
RingBurstISO    = 400
RingBurstShutterSpeed = (1/60)

-- ** TOTALITY ***************************************************************************
-- During the time between C2 and C3, the script will run back and forth between the 
-- "MinShutterSpeed" and "MaxShutterSpeed" as quickly as possible, with an extra 2 long exposures
-- at midpoint.  "ExpStep" is the size of the f-stop variation, and can be set to 0.333333, 1, 2, etc.
-- Min, Max and PrefISO:  Totality exposures will run (where possible) at PrefISO.  However, there
-- will also be exposures at MinShutterSpeed from MinISO to PrefISO, and MaxShutterSpeed from
-- PrefISO to MaxISO.
--
TotalityMinISO          = 100 
TotalityMaxISO          = 800
TotalityPrefISO         = 200
TotalityMinShutterSpeed = (1/250) -- (MinShutterSpeed is the *fastest* speed to use.)
TotalityMaxShutterSpeed = 1       -- 1 sec OK (MaxShutterSpeed is the *slowest*, longest speed used.)
TotalityExpStep         = 1  

-- ** ASHEN LIGHT **********************************************************************
-- One of the more difficult exposures to capture is earthshine -- the surface of the moon,
-- illuminated by light reflected from the earth.
--
-- The best time to do this is at the point of maximum eclipse, where the sun is centred behind
-- the moon, as much as possible.  Exposures here are kind of guesswork, and I have actually turned this
-- off by default.
--
DoMaxExposures  = 1		-- Number of (possibly bracketed) exposures to take at max-eclipse.
MaxOffset       = 3   -- How long before max eclipse to start these exposures.  You'll have to test
						          -- or use math to determine this value.  (Value is in seconds.)
NumMaxExposures = 1
DoMaxBrackets   = 1		-- Brackets?
NumMaxBrackets  = 1 
MaxBracketStep  = 1
MaxISO = 800
MaxShutterSpeed = 1

--
-- Optionally enable high speed burst mode
--
-- As originally written, the script used the camera.burst() function, to take a burst of
-- images as quickly as possible, as though the shutter button was held down.  This had two issues:
-- The first being that the buffer in the camera fills after around 2-3 seconds when shooting
-- RAW, and then slows drastically, and the second being that once called, the camera will take
-- exposures until the requested number of images are captured, and if the buffer fills, this may
-- take considerably longer than expected.
--
-- I have (by default) replaced this function with a burst function that tries to take a given number 
-- of images over a set period of time, each with a single shutter release call.  This is much slower 
-- (two or three frames per second, tops) but therefor fills the buffer more slowly, and allows the script
-- to abort the burst (and take fewer than the requested number of images) if the capture runs past the 
-- specified time limit.
--
-- If you prefer the old high speed burst call, set this variable to 1.
--
UseBurst = 0

-- ** END SETTINGS **************************************************************************************


-- ** START CALCULATIONS ***  BE CAREFUL BELOW THIS LINE ************************************************
--
-- Times are easiest to deal with in seconds.  This would be painful if they crossed over midnight,
-- but late-night solar eclipses are rare.
--
c1_sec = c1.hr * 3600 + c1.min * 60 + c1.sec 
c2_sec = c2.hr * 3600 + c2.min * 60 + c2.sec 
c3_sec = c3.hr * 3600 + c3.min * 60 + c3.sec 
c4_sec = c4.hr * 3600 + c4.min * 60 + c4.sec 
max_sec = math.floor(c2_sec + ((c3_sec - c2_sec) / 2))

MaxOffset = MaxOffset * DoMaxExposures -- This is ugly, and shouldn't be here.

tick_offset   = 0
TestStartTime = 0

--
-- Log to stdout and optionally to a file
--
function log (s, ...)
	local str = string.format (s, ...)
	str = str .. "\n"
	if (LogToFile == 0 or LoggingFile == nil)
	then
		io.write (str)
	else
		LoggingFile:write (str)
	end
	return
end

--
-- Open log file
--
function log_start ()
	if (LogToFile ~= 0)
	then
		local cur_time = dryos.date
		-- Opening logger with long filename fails. Works with short name.
		-- local filename = string.format("eclipse_%04d%02d%02d_%02d%02d%02d.log", cur_time.year, cur_time.month, cur_time.day, cur_time.hour, cur_time.min, cur_time.sec)
		-- local filename = string.format("EM%02d%02d.log", cur_time.hour, cur_time.min)
		local filename = string.format("eclipse.log")
		print (string.format ("Open log file %s", filename))
		LoggingFile = logger (filename)
	else
		print (string.format ("Logging not configured"))
	end
end

--
-- Close log file
--
function log_stop ()
	if (LogToFile ~= 0)
	then
		print (string.format ("Close log file"))
		LoggingFile:close ()
	end
end

--
-- Get the current time (in seconds) from the camera's clock.
--
function get_cur_secs ()

	local cur_time = dryos.date
	local cur_secs = (cur_time.hour * 3600 + cur_time.min * 60 + cur_time.sec)
	
	if ( TestBeepNoShutter == 1 )
	then
		cur_secs = (cur_secs - TestStartTime)	 -- If we're testing, start the clock at 
 													                 -- now, not actual time.
	end	
	return cur_secs

end

--
-- Take a time variable expressed in seconds (which is what all times are 
-- stored as) and convert it back to HH:MM:SS
--
function pretty_time (time_secs)

	local text_time = ""
	local hrs  = 0
	local mins = 0
	local secs = 0
	
	hrs  =  math.floor(time_secs / 3600)
  mins = math.floor((time_secs - (hrs * 3600)) / 60)
	secs = (time_secs - (hrs*3600) - (mins * 60))
	
	text_time = string.format("%02d:%02d:%02d", hrs, mins, secs)
	
	return text_time

end

--
-- Take a shutter speed expressed in (fractional) seconds and convert it to 1/x.
--
function pretty_shutter (shutter_speed)

	local text_time = ""
	if (shutter_speed >= 1.0)
	then
		text_time = tostring (shutter_speed)
	else
		text_time = string.format ("1/%s", tostring (1/shutter_speed))
	end
	return text_time

end

--
-- Hurry up and wait for the next important time to arrive.
--
-- Leave the console displayed for 60 seconds at the start and end of 
-- a wait.  Turn it off between, so that tracking can be done via live view, etc.
--
function wait_until (done_waiting)

	local counter = get_cur_secs()
	local next_sec = 0
	local show_console = ConsoleShowDelay
	local console_visible = 1
	
	console.show()
	
	log ("Waiting for %s in %d seconds.", pretty_time(done_waiting), done_waiting - counter)
	
	repeat

		task.yield (1000) -- Let the camera do other tasks for a second.
		
		if ((HideConsole == 1) and (show_console > 0))
		then
			show_console = show_console -1
		elseif ((HideConsole == 1) and (show_console == 0))
		then
			console.hide()
			show_console    = -1
			console_visible = 0
		end
		
		if ((HideConsole == 1) and ((done_waiting - counter) < 30 ) and (console_visible == 0))
		then
			console.show()
			console_visible = 1
		end
		
		counter = get_cur_secs()
					
	until (counter >= (done_waiting - 1))
			
	if ( counter < done_waiting)	
	then
									-- Loop /should/ exit the second before we are done. But
									--  It's possible that it could exit early in our target
									--  second. If so, we don't want to wait around to 
									--  (done_waiting + 1) to exit.
		next_sec = (1000 - ((dryos.ms_clock - tick_offset) % 1000))
		msleep (next_sec) -- Hard sleep, don't let anything else have priority.
	end
				
end

--
-- Set up the camera, and take a picture.  Also deals with any requested bracketing.
--
function take_shot(iso, shutter_speed, dobkt, bktstep, bktcount)

	local bktspeed = 0.0
	
	if ((lv.enabled == true) and (WarnLiveView == 1))
	then
		print ("TURN LIVEVIEW OFF (OR DISABLE MOVIE MODE) AND PRESS A BUTTON!!!")
		do_beep()
		key.wait()
	end
	
	camera.iso.value = iso
	
	if (dobkt == 0)
	then	-- Single exposure
		camera.shutter.value = shutter_speed
		log ("Click! Time: %s  ISO: %s  shutter: %s", 
			pretty_time(get_cur_secs()), tostring(camera.iso.value), pretty_shutter(camera.shutter.value))
		if (DoShutterShockDelay == 1)
		then
			if ((shutter_speed >= FastestDelayedShutter) and (shutter_speed 
				<= SlowestDelayedShutter))
			then
				task.yield(ShutterShockDelayMS)
			end
		end
	
		if (TestBeepNoShutter == 0) 
		then
			camera.shoot(false)
			task.yield(10) -- Exposures can take time.  Give other stuff a chance to run.
		else
			beep(1,50)
			task.yield (600 + camera.shutter.ms)
		end
		
	else	-- Bracketing exposure
		    -- Loop through the requested number of exposure brackets.
		for bktnum = bktcount,(-1 * bktcount),-1 do
			bktspeed = shutter_speed * (2.0^(bktnum * bktstep))
			camera.shutter.value = bktspeed
			log ("Click! Time: %s  ISO: %s  shutter: %s", 
				pretty_time(get_cur_secs()), tostring(camera.iso.value), pretty_shutter(camera.shutter.value))
			if (DoShutterShockDelay == 1)
			then
				if ((shutter_speed >= FastestDelayedShutter) and (shutter_speed 
					<= SlowestDelayedShutter))
				then
					task.yield(ShutterShockDelayMS)
				end
			end
			
			if (TestBeepNoShutter == 0) then
				camera.shoot(false)
				task.yield(10) -- Give other stuff a chance to run
			else
				beep(1,50)
				task.yield ((600 + camera.shutter.ms))
			end
		end
	end
end


--
-- Burst is simpler than single shot, because no brackets.  Set the camera and shoot.
--
-- I have tried replacing this with multiple calls to "camera.shoot()", but it is still considerably
-- slower than "camera.burst()", and can also crash the camera -- nothing permanent, but still
-- not something I want to put out.
--
function take_burst (count, iso, speed)

	camera.shutter.value = speed
	camera.iso.value = iso
	
	if ((lv.enabled == true) and (WarnLiveView == 1))
	then
		print ("TURN LIVEVIEW OFF (OR DISABLE MOVIE MODE) AND PRESS A BUTTON!!!")
		do_beep()
		key.wait()
	end
	
	log ("Burst! Time: %s  ISO: %s  shutter: %s  count: %d", 
		pretty_time(get_cur_secs()), tostring(camera.iso.value), pretty_shutter(camera.shutter.value), count)
	
	if (TestBeepNoShutter == 0)
	then
		camera.burst(count)
		task.yield(10)
	else
		beep(3,50)
		task.yield (4000 + (count * camera.shutter.ms))		
	end

end

--
-- Take X pictures over Y seconds
--
-- The camera.burst() function is useful for taking exposures as fast as possible, but is
-- limited by the camera's buffer space.  Depending on the body, burst mode will fill the buffer in 
-- around 2 seconds.  Even the slower cameras will fill the buffer with RAW images in 3-4 seconds, 
-- which is too fast to reliably capture Baily's Beads.  Switching to JPG would help, but must be 
-- done manually.  Not a good option for several reasons.
--
-- This function implements a manually controlled burst mode, which stretches out the exposure speed.
-- This spreads the time where the ~14 exposures that generally fit into the buffer out over a longer
-- time, and also gives the camera time to write to the card.  Instead of 14 frames over 4 seconds,
-- (~4fps) then slowing to maybe 4 frames every 3 seconds (1.5fps), we might be able to sustain 
-- 2fps for 10 seconds.  Actual best framerate and duration will need testing for each camera and
-- memory card.
--
-- Timing is more important than number of exposures, so this function will exit at the end of the
-- specificed timespan, even if the required number of images have not been taken yet.
function take_timed_burst(count, timespan, iso, speed)

	local start_time     = dryos.ms_clock -- Millisecond clock time that we're starting.
	local end_time       = (dryos.ms_clock + (timespan * 1000)) -- Clock time (in ms) where that we're done.
	local burst_interval = ((timespan * 1000) / count) -- Time between shutters, in milliseconds.
	local time_now       = start_time
	local pause_time     = 0
	local exposure_num   = 0
	local last_time      = 0
	
	for exposure_num = 1, count, 1
	do
		last_time =time_now
		time_now  = dryos.ms_clock
		take_shot (iso, speed, 0, 0, 0)
		if (time_now > end_time)
		then
			return -- No time for another image.
		else
			pause_time = ((start_time + burst_interval * exposure_num) - (dryos.ms_clock + 75))
			if (pause_time > 0) -- Pause for the next interval to pass, if we're not running late.
			then
				task.yield(pause_time)
			end
		end
		
	end

end

-- 
-- Simple camera beep.
--
function do_beep()  
		beep (5,100)
end
--
-- Take the spaced exposures for the C1-C2 and C3-C4 periods.  Take the margin times off either
-- end, split the time into the right intervals, and fire off take_picture()
--
function do_partial (start_phase, stop_phase, which_partial)

	local image_time = 0
	local image_interval = math.floor((stop_phase - start_phase) / (PartialExposureCount))
	local exposure_count = 0
	
	if (SetAperture == 1)
	then
		camera.aperture.value = PAperture -- Set aperture for partial phase shots
		log ("Aperture %s for partial phase shots", PAperture)	
	end	
	
	-- In a series of (PartialCount + 1) images, totality is either
	--	the first or last image.  This arranges the timing so that there
	--	will be a equidistance set of ((2 x PartialCount) + 1) exposures,
	--	with totality properly centered.
	
	if (which_partial == "Pre")
	then
		image_time = start_phase
	else
		image_time = start_phase + image_interval
	end
	
	if ( get_cur_secs() >= stop_phase ) -- Are we past this phase already?
	then
		log ("Skip %s Partial. Finished %d seconds ago.", which_partial, (get_cur_secs() - stop_phase))
		return
	end
	
	repeat
	
		log ("%s Partial: %d/%d  Interval: %d s  Remaining: %d",
			which_partial, exposure_count + 1, PartialExposureCount, image_interval, stop_phase - get_cur_secs())
		if (get_cur_secs() <= image_time)
		then
		
-- Reminder to center sun
			wait_until (image_time - 30)
			print()
			print("********************************************************")
			log  ("30 seconds to partial shot! Center!")
			print("********************************************************")
			print()
			do_beep()

-- Shoot		
			wait_until(image_time)
			take_shot (PartialISO, PartialShutterSpeed, PartialDoBkt, PartialBktStep, PartialBktCount)
			
		end
		
		image_time = image_time + image_interval
		exposure_count = exposure_count + 1
	    
-- Reminder at mid partial
		if ( exposure_count == math.floor(PartialExposureCount/2)+1 )
		then
  		  print()
		  print("********************************************************")
		  log  ("Mid partial phase. Check focus w. lunar limb!")
		  print("********************************************************")
		  print()		
		  do_beep()
		end
		
	until (exposure_count >= PartialExposureCount)
	
end

--
-- Start the burst shot a little before C2, then start running through exposure settings, going from
-- short, fast exposures to slow, long exposures and then back to short, until just before the midpoint
-- of the eclipse.  <strikethrough>Take two long exposures at that point, for good measure.</strikethrough>
--
function do_c2max()

	local cur_shutter_speed = 0
	local CurISO = 0
	
	if ( get_cur_secs() >= max_sec ) -- Are we past this phase already?
	then
		log ("Skip C2->Max. Finished %d seconds ago.", (get_cur_secs() - max_sec))
		return
	end
	
	if (get_cur_secs() <= (c2_sec - (math.max(RingStartOffset, C2BurstStartOffset) + 30)))
	then
		log ("Main C2->Max loop for %d seconds.", (c2_sec - (math.max(RingStartOffset, C2BurstStartOffset) + 30)))
		wait_until (c2_sec - (math.max(RingStartOffset, C2BurstStartOffset) + 30))
		print()
		print("********************************************************")
		log  ("30 seconds to C2! Remove Filter! Center! ")
		print("********************************************************")
		print()
		do_beep()
	end
	
	if (SetAperture == 1)
	then
		camera.aperture.value = C23Aperture -- Set aperture for C2 shots
		log ("Aperture %s for C2 shots", C23Aperture)
	end	
	
	if ((get_cur_secs() <= (c2_sec - RingStartOffset) ) and ( DoRing == 1)) -- Are we taking Diamond Ring shots?
	then
		wait_until (c2_sec - RingStartOffset)
		if (UseBurst == 1)
		then
			take_burst (RingBurstCount, RingBurstISO, RingBurstShutterSpeed)
		else
			take_timed_burst (RingBurstCount, RingBurstTime, RingBurstISO, RingBurstShutterSpeed)
		end
	end
	
	if ( get_cur_secs() < c2_sec ) -- Have we passed the burst for Baily's beads?
	then
		wait_until (c2_sec - C2BurstStartOffset)
		if (UseBurst == 1)
		then
			take_burst (C23BurstCount, C23BurstISO, C23BurstShutterSpeed)
		else
			take_timed_burst (C23BurstCount, C2BurstTime, C23BurstISO, C23BurstShutterSpeed)
		end
		print()
		print("********************************************************")
		log  ("Post C2 warning!")
		print("********************************************************")
		print()		
		do_beep()
		
	end
	
	cur_shutter_speed = TotalityMinShutterSpeed
	CurISO = TotalityMinISO
	if (SetAperture == 1)
	then
		camera.aperture.value = Aperture -- Reset aparture for totality shots
		log ("Aperture %s for totality", Aperture)	
	end	
	
	repeat
	
		take_shot(CurISO, cur_shutter_speed, 0, 0, 0)

		if (CurISO < (TotalityPrefISO * 0.95))
		then
			CurISO = CurISO * 2.0^TotalityExpStep
		elseif ((CurISO < (TotalityPrefISO * 1.1)) and (cur_shutter_speed < (TotalityMaxShutterSpeed * 0.95)))
		then
			cur_shutter_speed = cur_shutter_speed * 2.0^TotalityExpStep
		elseif (CurISO < (TotalityMaxISO * 0.95))
		then
			CurISO = CurISO * 2.0^TotalityExpStep
		else
			cur_shutter_speed = TotalityMinShutterSpeed
			CurISO = TotalityMinISO
		end
			
	until (get_cur_secs() >= (max_sec - MaxOffset)) -- Stop, and leave time to do the mid-eclipse earthshine
													-- exposures.
end

--
-- do_max -- Take a number of long exposures at the time of maximum eclipse, to try to capture
-- an earthshine image.  These can be bracketed.
--
function do_max()

	if (DoMaxExposures == 0)  -- Are we doing this?
	then
		log ("Skip Max Eclipse long exposures. Not configured.")
		return 		-- Nope.
	end

	if ( get_cur_secs() >= max_sec) -- Have we passed max already?
	then
		log ("Skip Max. Finished %d seconds ago.", (get_cur_secs() - max_sec))
		return
	end

	for count_max_exp = NumMaxExposures , 1 , -1 do
		log ("do_max: MaxISO=%d, MaxShutterSpeed=%s, DoMaxBrackets=%d, NumMaxBrackets=%d, MaxBracketStep=%d",
			MaxISO, tostring(MaxShutterSpeed), DoMaxBrackets, NumMaxBrackets, MaxBracketStep)
		take_shot(MaxISO, MaxShutterSpeed, DoMaxBrackets, MaxBracketStep, NumMaxBrackets)
	end
	
end
 
--
-- Similar to do_c2max, but reversed.  Exposures run from longest to shortest (and then repeat), 
-- the burst starts just before C3, and there are no bonus exposures.
--
function do_maxc3()

	local cur_shutter_speed = 0
	local CurISO = 0

	if ( get_cur_secs() >= (c3_sec + RingStartOffset) ) -- Are we past this phase already?
	then
		log ("Skip Max->C3. Finished %d seconds ago.", (get_cur_secs() - (c3_sec + C3BurstStartOffset)))
		return
	elseif ( get_cur_secs() < (c3_sec - C3BurstStartOffset) ) -- Do we have time for some totality exposures?
	then
		cur_shutter_speed = TotalityMaxShutterSpeed
		CurISO = TotalityMaxISO
		log ("Main Max->C3 loop for %d seconds.", (c3_sec - C3BurstStartOffset - get_cur_secs()))
		repeat
			take_shot(CurISO, cur_shutter_speed, 0, 0, 0)
			if (CurISO > (TotalityPrefISO * 1.05))
			then
				CurISO = CurISO / 2.0^TotalityExpStep
			elseif ((CurISO > (TotalityPrefISO * 0.95)) and (cur_shutter_speed > (TotalityMinShutterSpeed * 1.05)))
			then
				cur_shutter_speed = cur_shutter_speed / 2.0^TotalityExpStep
			elseif (CurISO > (TotalityMinISO * 1.01))
			then
				CurISO = CurISO / 2.0^TotalityExpStep
			else
				cur_shutter_speed = TotalityMaxShutterSpeed
				CurISO = TotalityMaxISO
			end
						
		until (get_cur_secs() >= (c3_sec - (C3BurstStartOffset + 3)))

		print()
		print("********************************************************")
		log  ("3 seconds to C3!  Filter warning!")
		print("********************************************************")
		print()
		do_beep()
			
	end
	
	if (SetAperture == 1)
	then
		camera.aperture.value = C23Aperture -- Set aperture for C3
		log ("Aperture %s for C3 shots", C23Aperture)
	end	
	
	wait_until (c3_sec - C3BurstStartOffset)
	
	if (UseBurst == 1)
	then
		take_burst (C23BurstCount, C23BurstISO, C23BurstShutterSpeed)
	else
		take_timed_burst(C23BurstCount, C3BurstTime, C23BurstISO, C23BurstShutterSpeed)
	end
	
	if (DoRing == 1)
	then
		if (UseBurst == 1)
		then
			take_burst (RingBurstCount, RingBurstISO, RingBurstShutterSpeed)
		else
			take_timed_burst(RingBurstCount, RingBurstTime, RingBurstISO, RingBurstShutterSpeed)
		end
	end

	print()
	print("*************************************************************")
	log  ("End of totality! Replace filter! Display Off!")
	print("*************************************************************")
	print()
	do_beep()

end

--
-- The ringleader.
--
function main()

	local starttime
	local offset = 0
	local offset_count = 0

	starttime = get_cur_secs()
	TestStartTime = starttime
	
    menu.close()
    console.show()
	log_start ()

	--
	-- The camera maintains a millisecond timer since power-on.  We can use this to
	-- get close to the beginning of a given second.  I think.
	--
	event.seconds_clock = function (ignore)
	
	offset = offset + (dryos.ms_clock - (1000 * offset_count))
	offset_count = offset_count + 1
			
	return true
	
end
	print ()
	print ()
	print ("-------------------------------------")
	print ("  Eclipse Magic")
	print ("  Copyright 2017, grnbrg@grnbrg.org")
	print ("  Mod 2019-2020 by G Abramson")
	print ("  Released under the GNU GPL")
	print ("-------------------------------------")
	print ()
	print ("Starting 10 second timing calibration....")

	--
	-- There is a fair amount of jitter in the event timer.  Averaging over 10 seconds will
	-- give us a reasonable offset.
	--
	task.yield(10500)

	-- Turn off the second_clock event timer.
	event.seconds_clock = nil
	
	tick_offset = (math.floor(offset / offset_count) % 1000)
	
	print ("Done!")
	print ()
	log ("TestBeepNoShutter: %d", TestBeepNoShutter)
	log ("C1: %s", pretty_time(c1_sec))
	log ("C2: %s", pretty_time(c2_sec))
	log ("C3: %s", pretty_time(c3_sec))
	log ("C4: %s", pretty_time(c4_sec))

	-- If the camera is not in manual mode, trying to set the shutter speed throws errors.
	-- Check to make sure we are in manual mode, and refuse to run if we're not.
	if (camera.mode == MODE.M)
	then
--		if (SetAperture == 1)
--		then
--			camera.aperture.value = Aperture
--           log ("Aperture %s", Aperture)			
--		end
	
		do_partial ((c1_sec + PartialMarginTime), c2_sec, "Pre")
	
		do_c2max()
		
		do_max()
	
		do_maxc3()
	
		do_partial (c3_sec, (c4_sec - PartialMarginTime), "Post")
		
	else
		beep (5, 100)
		log  ("Camera must be in manual (M) mode!!")
		print()
		print("Press any button to exit the script.  Change the mode and re-run.")
		key.wait()
	end
		
	log ("All done. Normal exit.")
	log_stop ()
	print("Press any button to exit the script.")
	key.wait()
    console.hide()
	
end -- Done.  Hope there were no clouds.

main() -- Run the program.

-- CHANGES

-- 1.0.1
	-- Stopped running the seconds_clock event at all times, and moved the tick_offset
		 -- calculation to the setup at the start of execution
	-- Fixed the pretty-printing of the timer
	-- Fixed the sign of the C23BurstStartOffset in do_maxc3
	-- Fixed the calculation of the difference between current time and the next second
		 -- in wait_until()
	-- Added code to turn off the console during long waits
	-- Massaged the end-of-exposure-bracketing conditions in do_c2max() and do_maxc3()
	
-- 1.1
	-- Changed the partial phase exposure logic, so that instead of there being an exposure
		 -- at C1 and C2, there is an exposure at C1 and an exposure at (C2 - exposure_interval)
		 -- so that the totality images are properly centered between the requested partial phase
		 -- exposures.
	-- Added an alarm at 30 seconds before C2 and 3 seconds before C3 to alert for any filter changes.  
		 -- There is also a beep after the C2 and C3 bursts to flag any needed changes.
	-- Split the C23BurstStartOffset into separate variables, to allow an asymmetric burst over
		 -- each period.  (ie: 10 seconds before C2 to 3 seconds after C2)
		
-- 1.2
	-- Added ISO brackets to totality exposure sequence.  Totality now has a preferred ISO, 
		 -- and will shift to that ISO at the fastest or slowest shutter speeds, then use that
		 -- preferred ISO for the requested range of shutter speeds, then shift ISO to the end of
		 -- the requested range.
	-- Option to stop LiveView before touching the shutter controls.
	-- Removed the two long exposures at max eclipse -- probably not needed.
	
-- 1.2.1
	-- Changed call to lv.stop() to a beep, and instruction to the user to turn off LV.
		 -- lv.stop() doesn't seem to work.
	-- Added print statements to explain filter warning beeps.
	
-- 1.3.0
	-- Changed the startup timing loop to be more accurate if the script is started close to
	   -- a second boundary.  (The average offset of 999ms and 1ms is 1ms, not 500ms)
	-- Corrected an error in wait_until() -- Used div, where modulus was correct, and had
	   -- the tick_offset correction wrong.  Don't code while tired.  Thanks to
	   -- matman730 for pointing out this goof.
	-- Improved the configuration comments around TestBeepNoShutter.  They apparently weren't
	   -- clear as I thought.
		
-- 1.4.0  (Not released)
	-- Attempt to implment the changing of the file prefix for the saved images.  It didn't
	   -- work well, and I scrapped it.
		
-- 1.5.0
	-- Make it optional to hide the console during script running, and allow the delay before
	   -- and after the next image to be configured.
	-- Shutter shock reduction:  Add a configurable (in milliseconds) delay before an exposure
	   -- where the shutter speed is within a (also configurable) range.
	-- Add a mid-eclipse section.  This is a short section around the max eclipse point to optionally
	   -- try for some earthshine exposures.
	-- Set the aperture on program start.
	
-- 1.6.0 -- Contributions from Eric Krohn, <krohn@ekrohn.com> (Many thanks!)
	-- Added logging to permanent file, "ECLIPSE.LOG" at top level of the memory card
	-- More extensive logging added throughout the script
	-- Bugfix:  In do_max() the last two arguments to take_shot() were reversed. 
	-- Bugfix:  do_max() is unguarded as far as current time
	-- Added pretty_shutter() to make the shutter speed numbers more sensible

-- 1.7.0
	-- Added optional function to take some diamond ring images before and after the Baily's Beads
	   -- exposures.  Be careful not to overlap the Ring and Beads exposures before C2!
	-- Added a new burst function that takes a burst of images, one at a time, rather than using the
	   -- camera's burst function.  Slower, but slower is better for the buffer, and gives us the
	   -- opportunity to stop shooting at a specific time, where a burst will run until the requested
	   -- number of images have been captured.

-- 1.7.1: 
    -- Added reminder at mid partial to check focus.
    -- Added reminder 30 sec before each partial shot to recenter.
	  -- Function beep: just one instead of 3 5-beeps.
	  -- Added apertures for partial, ring and beads, and totality.
\end{verbatim}

\end{document}